\documentclass[aps,groupedaddress,nofootinbib,11pt]{revtex4}

\usepackage[dvips]{epsfig}

\usepackage{amsfonts}

\usepackage{amsmath}

\usepackage{}

\usepackage{color}

\definecolor{Blue}{rgb}{0.3,0.3,0.9}

\definecolor{Red}{rgb}{1.0,0.0,0.0}

\definecolor{Green}{rgb}{0,0.4,0}

\definecolor{Violet}{rgb}{0.4,0.0,0.6}

\definecolor{Cyan}{rgb}{0.0,0.4,0.6}

\definecolor{Orange}{rgb}{1.0,0.4,0.0}

\newcommand{\Eqref}[1]{(\ref{#1})}

\newcommand{\ket}[1] {\mbox{$ \vert #1 \rangle $}}

\newcommand{\bra}[1] {\mbox{$ \langle #1 \vert $}}

\newcommand{\av}[1]{\langle\!\langle \,  #1 \, \rangle\!\rangle}

\newcounter{subequation}[equation] \makeatletter

\expandafter\let\expandafter\reset@font\csname
reset@font\endcsname

\newenvironment{subeqnarray}
  {\arraycolsep1pt
    \def\@eqnnum\stepcounter##1{\stepcounter{subequation}{\reset@font\rm
      (\theequation\alph{subequation})}}\eqnarray}
  {\endeqnarray\stepcounter{equation}}
\makeatother

\newcommand{\ba}{\begin{eqnarray}}
\newcommand{\ea}{\end{eqnarray}}
\newcommand{\sba}{\begin{subeqnarray}}
\newcommand{\sea}{\end{subeqnarray}}

\begin{document}

\vskip 1truecm

 \title{Damped Corrections to Inflationary Spectra from a Fluctuating Cutoff} 
\vskip 1truecm
 \author{David Campo and Jens Niemeyer}
 \email[]{dcampo@astro.uni-wuerzburg.de}
 \email[]{niemeyer@astro.uni-wuerzburg.de}
 \affiliation{Lehrstuhl f\"{u}r Astronomie,
Universit\"{a}t W\"{u}rzburg, Am Hubland,
D-97074 W\"{u}rzburg, Germany}
 \author{Renaud Parentani}
 \email[]{Renaud.Parentani@th.u-psud.fr}
 \affiliation{Laboratoitre
de Physique Th\'{e}orique, CNRS UMR 8627, B\^atiment 210,
Universit\'{e} Paris XI, 91405 Orsay Cedex, France}

 \begin{abstract}  
We reconsider trans-Planckian corrections
to inflationary spectra by taking into account 
a physical effect which has been overlooked
and which could have important consequences.
We assume that the short length scale 
characterizing the new physics 
is endowed with a finite width, the origin of which 
could be found in quantum gravity.
As a result, the leading corrections responsible for 
superimposed osillations in the CMB temperature anisotropies 
are  generically damped by the blurring of the UV scale. 
To determine the observational ramifications of this damping,
we compare it to that which effectively occurs when computing
the angular power spectrum of temperature anisotropies. 
The former gives an overall change of the oscillation amplitudes
whereas the latter 
depends on the angular scale.
Therefore, in principle they could be distinguished. 
In any case, the observation of superimposed oscillations would place a
tight constraint on the variance of the UV cutoff.  
 \end{abstract}

\maketitle

\section{Introduction}

In light of the impressive agreement of all current cosmological
observations with the paradigm of inflation and the generation of primordial
perturbations from quantum fluctuations \cite{WMAP3,MartinReview},
every opportunity for finding signs of new physics in the data should
be explored. Simple phenomenological models for new high
energy physics have recently been used in order to characterize
deviations from the standard predictions. This is the general approach
that we pursue here, analyzing 
an important physical effect that has
so far been overlooked.

Standard inflationary spectra are governed by $H$, the Hubble scale
during inflation, and its behavior as a function of the background
energy-momentum content (i.e. the inflaton potential in the simplest
scenarios). On the other hand, deviations may depend on a second scale
such as, for instance, the cutoff $M$ at which the standard low-energy
theory breaks down. To preserve the leading behavior, the new scale is taken to 
be much higher than other physical scales, i.e., here $H/M \ll 1$.
This line of thought was first applied to black hole radiation
\cite{Tedcutoff} and then transposed to the cosmological context in
\cite{MB00N00}. In both cases, when considering backward in time propagation,
the tremendous blueshift experienced by the mode frequency 
acts like a space-time microscope which brings the (proper) frequency
across the new scale \cite{TedRiver}. 
However, the adiabatic evolution of the quantum
state reduces the deviations of the outcoming spectra.
In inflationary cosmology, their  amplitude is proportional to a
positive power of $H/M$, which makes their detection very challenging.
\footnote{Note
  that the WMAP data has been reported to show marginal evidence for the 
  presence of oscillations in the power spectrum that may be explained
  by trans-Planckian effects \cite{WMAP3,MR}.}

In the present paper we extend previous analysis
by pointing out that it is unlikely that the $UV$ scale $M$  
be fixed with an infinite precision. 
On the contrary,
it is possible that the gradual appearance of new physics effectively
endows $M$ with a finite width. Whether this width arises from quantum
mechanics or from a classical stochastic process will be left
unspecified in this work; we will simply treat $M$ as a random
(Gaussian) variable and assume that its fluctuations are small with
respect to the mean. As expected, the average over the fluctuations
washes out all oscillatory corrections to the power spectra which depend on a 
rapidly varying phase. 
This is important because the leading corrections to the power
spectrum from a high-energy cutoff found so far, 
see e.g. \cite{MB03} and references therein, 
 are precisely functions of this type.

A similar damping mechanism was found in \cite{BFP} 
when considering the modifications of Hawking radiation 
induced by metric fluctuations treated stochastically. 
Furthermore, it was shown \cite{beyond} that the stochastic treatment
emerges from a quantum mechanical analysis of gravitational
loop corrections. This indicates that the phenomenology of blurring
the $UV$ scale is insensitive to the particular underlying mechanism.
We will demonstrate, however, that it can in principle be distinguished
from an adiabatic suppression since it only acts on
the oscillatory corrections, whereas the latter also affects the
slowly varying contributions. Hence, it opens the door to investigate
a new aspect of cutoff phenomenology with possible links to quantum gravity. 
Other phenomenological signatures of a fluctuating geometry have 
been considered in \cite{bibliofluct}.

To implement the notion of a fluctuating cutoff, we first use a 
phenomenological description in which each independent field mode of
wave number $q$ is placed into an instantaneous vacuum state at the time 
its redshifted momentum crosses $M$. Depending on the adiabaticity of
the state, the resulting modifications are more or less suppressed but  
the leading correction is {\it always} a rapidly oscillating function of $M$
(and of $q$ in slow roll inflation). 
Therefore, in this class of models, the effect of averaging over the
fluctuations of $M$ damps the leading correction. The damping factor
depends on the width of $M$, but the crucial fact is that a
tiny variance (in units of the mean $\bar M$) is enough to eradicate
the oscillatory modifications of the power spectrum because their
frequency is very high (proportional to $M/H \gg 1$).

The paper is organized as follows. 
In Sec.\ \ref{sec:steady-non-steady} we summarize the derivation of the power
spectrum modifications and explain why they can 
be decomposed into a rapidly oscillating and a steady part. While this 
conclusion is reached for a particular class of models, in Sec.\
\ref{sec:gener-other-class} we generalize it to a 
wider class of possible modifications of the power spectrum. 
The process of averaging over stochastic fluctuations of the cutoff is
carried out in Sec.\ \ref{sec:stochastic-averaging}. 
We then point out that the UV-blurring shows some similarities with
the averaging involved in computing the multipole coefficients of the 
Cosmic Microwave Background temperature anisotropies 
from the primordial spectrum. We compare these effects in Sec.\
\ref{sec:geometric-average} and discuss our results in Sec.\
\ref{sec:discussion}.

\section{Steady and oscillatory corrections to power spectrum}

\label{sec:steady-non-steady}

We begin with a summary of the phenomenological description of 
trans-Planckian signatures arising from the choice of the initial
state of the modes of linear perturbations.  
The various elements are presented with the aim to highlight  
the origin and properties of the deviations from the standard
power spectrum. This presentation generalizes that of \cite{Easther} in that 
we derive the oscillatory properties of the leading correction 
in a wider context, and explain the origin of their universal character.

In inflationary models with one inflaton, the power spectra of both
linear curvature $\zeta$ and gravitational waves  $h_{ij}$ during inflation 
can be related to that of a quantum massless test field $\varphi$ as follows.
The scalar and tensor perturbations parameterized by $\zeta$ and 
$h_{ij}$ can be defined conveniently in the coordinate
system in which the inflaton field is homogeneous on 
spatial hypersurfaces, i.e. 
\begin{eqnarray}
  ds^2 &=& - N^2 dt^2 + \gamma_{ij} \left( dx^i + N^i dt  \right)  \,  
  \left( dx^i + N^i dt  \right) \, , 
\nonumber \\ 
  \delta \phi &=& 0 \, , \qquad  
 \gamma_{ij} = a^2(t) \left\{ \, \left( 1+2\zeta \right) \delta_{ij} +
   h_{ij} \right\}  
 \, , \qquad h_{\,\, i}^{i} = 0 \, \quad \partial_i h_{ij}= 0\, .
\end{eqnarray} 
The advantage of this gauge is that the metric perturbations are
physical degrees of freedom, and $\zeta$ has the remarkable property
of being constant outside the horizon \cite{constantzeta}.
Solving for the momentum and Hamiltonian constraints one obtains
\begin{eqnarray}
  N = 1+\frac{\partial_t \zeta}{H} \, , \qquad 
  N_i = \partial_i \left( -\frac{\zeta}{a^2H} + \epsilon_1 \nabla^{-2} 
  \partial_t \zeta  \right) \, , 
\end{eqnarray} 
where 
\begin{eqnarray}
  \epsilon_1 = - \frac{d \ln H}{d\ln a} = -\frac{\partial_t H}{H^2}\, . 
\end{eqnarray}
After introducing the auxiliary scalar field $\varphi$, the power
spectra of $\zeta$ and gravitational waves are obtained from 
that of $\varphi$ by the substitutions \cite{MukhaPhysRep} 
\begin{eqnarray} \label{substitution}
  \zeta = \varphi \, \frac{\sqrt{4\pi G}}{a\sqrt{\epsilon_1}}
   \, , \qquad 
   h_{ij} = \varphi \, \frac{\pi^{s}_{ij}}{a}\, , 
\end{eqnarray}
where $\pi^{s}_{ij}$ is the polarization tensor of the gravitational waves.
Given this correspondence, it is sufficient to understand the behavior
of $\varphi$.

Let us consider that each mode of $\varphi$ is imposed to be in a
given vacuum state $\ket{\Psi_M}$ at the time $t_M(q)$ when 
\begin{eqnarray} \label{t_M}
  q = M a(t_M) ,
\end{eqnarray}
that is, when the physical momentum $q/a$ crosses the proper scale $M$.
In this case, the power spectrum ${\cal P}_{M}(q)$ is 
related to the Fourier transform of the equal time two-point function
evaluated in $\ket{\Psi_M}$ by
\begin{eqnarray} \label{powervarphi}
  \bra{\Psi_M} \hat \varphi(t,{\bf x}) \hat \varphi(t,{\bf y}) \ket{\Psi_M} = 
 \int_{0}^{+\infty}\frac{dq}{q} \, \frac{\sin(qr)}{qr} \, {\cal P}_{M}(q,t)\, , 
 \label{Pws}
\end{eqnarray}
where $r = \vert {\bf x} - {\bf y} \vert $, 
and where the time $t$ is taken to be several e-foldings after
$t_H(q)$, the time of Hubble scale crossing for the mode $q$:
\begin{eqnarray}
    q= H(t_H) \, a(t_H) \, .
\end{eqnarray} 

In this paper, we assume that the Hubble scale is well separated
from the UV scale $M$, hence
\begin{eqnarray} \label{sigmaq}
   \sigma_q \equiv \frac{H_q}{M} \ll 1 \, ,
\end{eqnarray}
where $H_q$ is the value of $H$ evaluated at $t_H(q)$, 
see Figure 1.

\begin{figure}
\includegraphics[width=0.8\textwidth]{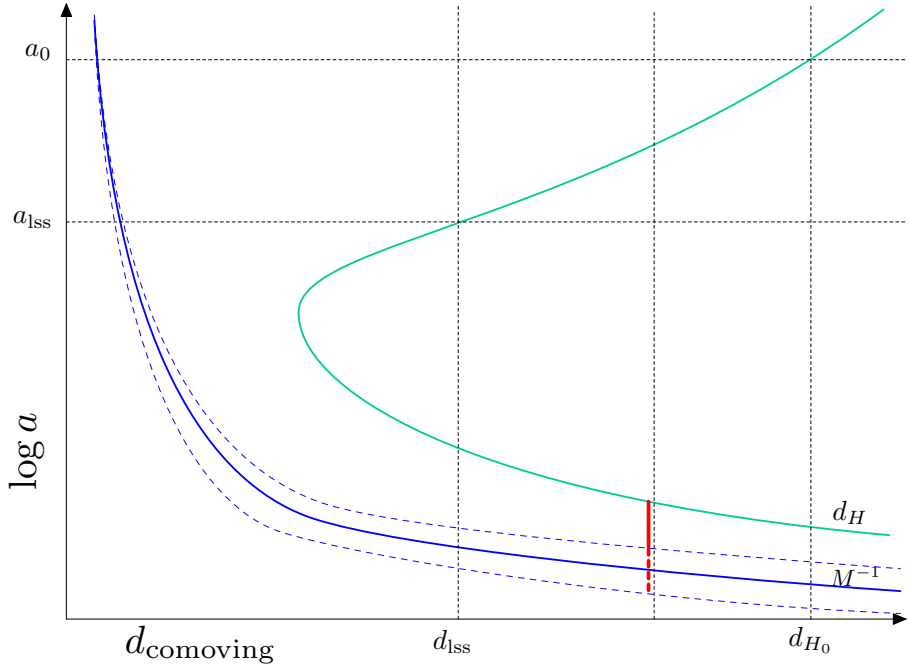}
\caption{Evolution of 
the comoving Hubble radius, $d_H= 1/Ha$, as a function of $\ln a$
during inflation (decreasing $d_H$) and radiation domination (growing
$d_H$), compared to the high energy comoving scale $1/Ma$. 
The two dotted lines represent the spread $\pm \Sigma/M^2$ about the mean. 
The vertical lines correspond to
the comoving scales of the Hubble radius today, at last scattering,
and at an intermediate time.
During slow roll inflation, the lapse of time between $t_M$ ($M$-crossing)
and $t_q$ ($H$-crossing, or horizon exit) increases as $d=1/q$ decreases,
giving rise to the $q$-dependence of $\sigma$, see Eq.\ \Eqref{sigmaq}.
The thick vertical line between the comoving curves represents the 
accumulated phase of the corresponding mode.}
\end{figure}

The definition of the vacuum state $\ket{\Psi_M}$
and the value of the power spectrum ${\cal P}_{M}(q)$ are both  
given in terms of the corresponding family of positive frequency
solutions (hereafter called  $\varphi_q^{M}$) of the mode equation 
\begin{eqnarray} \label{modeeqn}
  \left( \partial_{\tau}^2 + \omega_q^2(\tau) \right) \varphi_q = 0 \, .
\end{eqnarray}
Here, $\tau$ is the conformal time defined by $d\tau = dt/a(t)$ and
$\omega_q$ the conformal frequency whose properties will be
discussed below. The initial state $\ket{\Psi_M}$ is defined as the state 
annihilated by the destruction operators $\hat a_{\bf q}^M$ associated
with the modes $\varphi_q^{M}$.  These operators are given by the
Klein-Gordon overlap with the field operator $\hat \varphi$
\begin{eqnarray} 
  \hat a_{\bf q}^M =  \, \varphi_q^{M\, *} \,
  \overleftrightarrow{i \partial}_{\!\!\tau} \, \left(\int_{\tau =
      {\rm cte}}\!\!d^3x \,  
  \frac{e^{-i{\bf q}{\bf x}}}{(2\pi)^{3/2}} \, \hat \varphi (\tau, {\bf x})
  \right)\, .
\end{eqnarray}
 Straightforward algebra gives the power spectrum of Eq.\ (\ref{Pws}):
\begin{eqnarray} \label{generalpower}
  {\cal P}_M(q,t) = \frac{q^3}{2\pi^2}  \vert  \varphi_q^{M} (t) \vert^2 \, .
\end{eqnarray}
As in any vacuum state, it is given by the square of the norm of the
corresponding positive frequency modes evaluated long after horizon
crossing.

The standard spectrum also belongs to this class. It is obtained when using
the asymptotic vacuum, often called the Bunch-Davis vacuum 
\cite{BunchDavis}. 
This state is defined by the solutions of Eq.\ (\ref{modeeqn}) 
with positive frequency in the asymptotic past. 
Using the fact that $\omega_q \to q $ for $\tau \to -\infty$ (see
Eq.\ \ref{omegasl} below), the asymptotic positive frequency modes obey
\begin{eqnarray} \label{asympmodep}
  \left( i\partial_{\tau} - q \right)\varphi_q^{-\infty} 
   \vert_{\tau \to -\infty} = 0 \, .
\end{eqnarray}
The corresponding power spectrum is thus
\begin{eqnarray} \label{stdpower}
 {\cal P}_{-\infty}(q,t) &=& \frac{q^3}{2\pi^2}  \vert  \varphi_q^{-\infty}(t) 
  \vert^2 \, .
\end{eqnarray}

In the long wavelength limit, when $t \gg t_H(q)$, the 
standard spectra of the metric perturbations  
obtained using \Eqref{substitution} become constant and
depend only on $H_q$ and the hierarchy of slow roll parameters $\epsilon_n$
which are logarithmic derivatives, $\epsilon_1 = -d \ln H /d{\ln a}$ and 
$\epsilon_{n\geq 2} = d \ln \vert \epsilon_{n-1} \vert /d{\ln a}$
(we adopt the definition of \cite{beyondSL} in terms of the 
logarithmic derivatives of $H$ instead of the logarithmic derivatives of 
the inflaton potential). In the slow-roll approximation, $\epsilon_1$ and 
$\epsilon_2$ are constants and $\epsilon_{n\geq 3} = 0$.  In addition, the 
long wavelength limit of \Eqref{stdpower} is expanded to linear order in
$\epsilon_1,\epsilon_2$. Explicitly, the gravitational wave and
curvature spectra are given by (more details can be found in,
e.g. \cite{MS1}) 
\begin{eqnarray}
  P_{-\infty}^{\rm GW}  &=&  
  \frac{16 H^2}{\pi M_{\rm Pl}^2} \left[ 
  1 - 2(C+1)\epsilon_1   - 2 \epsilon_1 \ln\left( \frac{q}{q_0}  \right)
 \right] \, , \nonumber\\ 
 P_{-\infty}^{\zeta} &=& \frac{1}{\epsilon_1} \frac{H^2}{\pi  M_{\rm Pl}^2} \left[ 
  1 - 2(C+1)\epsilon_1 - C \epsilon_2  - 
 (2 \epsilon_1 + \epsilon_2)\ln\left( \frac{q}{q_0}  \right)
 \right]\, , 
\end{eqnarray} 
where $C=\gamma_E + \ln 2 - 2 \simeq -0.7296$,  $q_0$ is the
pivot-scale around which the expansion in $\ln(q)$ is carried out, and
the values of  $H, \epsilon_1, \epsilon_2$ are taken when $q_0$
crosses the horizon.

Since the modes 
$\varphi_q^{-\infty}$ and $\varphi_q^{M}$ obey the same equation, 
they are related by a time-independent transformation 
\begin{eqnarray} \label{newmode}
  \varphi_q^{M}(\tau) = \alpha_q \, \varphi_{q}^{-\infty}(\tau) + \beta_q \, 
  \varphi_{q}^{-\infty\, *}(\tau)
  \, .
\end{eqnarray}
As usual, the  Bogoliubov coefficients $\alpha_q$ and $\beta_q$ 
are given by the overlaps  
of the two sets of modes
\begin{eqnarray} \label{alphabeta}
   \alpha_q =   \, \left(\varphi_q^{-\infty}\right)^* \, 
   \overleftrightarrow{i\partial}_{\!\!\tau}   \,  \varphi_q^{M}   
 \, , \qquad 
  \beta_q =  - \varphi_q^{-\infty} \, 
   \overleftrightarrow{i\partial}_{\!\!\tau}   \,  \varphi_q^{M}\, .
\end{eqnarray}
Using these coefficients and \Eqref{stdpower} in the long wavelength limit,  
the power spectrum \Eqref{generalpower} is 
\begin{eqnarray} \label{newpower}
  {\cal P}_M(q) 
  &=& {\cal P}_{-\infty}(q) \, \times \vert \alpha_q \vert^2 
 \left\{  1 + 2 {\rm Re}\left( \frac{\beta_q^*}{\alpha_q^*} 
  \frac{\left(\varphi_{q}^{-\infty}\right)^2}{ \vert
    \varphi_q^{-\infty}  \vert^2}   
\right) + \frac{\vert  \beta_q\vert^2  }{\vert  \alpha_q\vert^2  } 
\right\}   \, .
\end{eqnarray}
This equation holds whenever new physics expresses itself
through the replacement of the asymptotic vacuum with a new vacuum state.
(It also applies for modified mode equations 
\Eqref{omegasl} including dispersion above $M$ \cite{np1}, 
see also \cite{MB03}.)

At this point, an important remark must be made. In Eq.\ \Eqref{newpower}, the
second term in the brackets is independent of the phase conventions
of the modes $\varphi_q^{-\infty}$ and $\varphi_q^{M}$.
Indeed, a change $\varphi^{-\infty} \mapsto e^{i\rho} \varphi^{-\infty}$ and 
$\varphi^{M} \mapsto e^{i\sigma} \varphi^{M}$ gives 
$\beta \mapsto e^{i(\rho + \sigma)} \beta$ and 
$\alpha \mapsto e^{i(\sigma - \rho)} \alpha$, from which follows the 
invariance of $\alpha \beta^* \left(\varphi^{-\infty}\right)^2$.
In other words, the phase of this term is physically 
meaningful. Moreover, it will play a key role in the averaging process
discussed below.

The corrections in \Eqref{newpower}, whose properties will now be
explained, result from the fact that the vacuum $\ket{\Psi_M}$
is less adiabatic than the asymptotic vacuum. Consider, for instance,
positive frequency modes obeying
\begin{eqnarray} \label{Mmode}
  \left( i\partial_{\tau} - 
  \omega_q(\tau)  \right)\varphi_q^{M} = 0 \, ,
\end{eqnarray}
at the time $\tau_M(q)$ defined by Eq.\ (\ref{t_M}). To characterize the
degree of adiabaticity, we specialize to slow-roll inflation.
In this case, the conformal frequency of metric perturbations is of the form
\begin{eqnarray} \label{omegasl}
  \omega_q^2(\tau) = q^2 - \frac{f}{\tau^2} \, 
\end{eqnarray}
where $f= f (\epsilon_1, \epsilon_2)$ is a constant 
of order unity whose explicit expression is not needed here. In the
above equation, we have chosen the arbitrary additive
constant in conformal time $\tau$ such that 
\begin{eqnarray} \label{slowrolltau}
  aH \simeq -\frac{(1+\epsilon_1)}{\tau} \, .
\end{eqnarray}

The degree of adiabaticity of the modes is controlled by the ratio \cite{np1}
\begin{eqnarray}
  \frac{\vert \partial_{\tau} \omega_q \vert}{\omega_q^2} = 
  \frac{f}{\vert q^2\tau^2 - f \vert^{3/2}} \, .
\end{eqnarray}
Applied to \Eqref{omegasl}, it shows that the evolution 
is adiabatic for $q \vert \tau \vert \gg 1$, i.e. when the wavelength
is much smaller than the Hubble radius $H^{-1}$.
Furthermore, the asymptotic evolution is WKB exact, i.e. there is no
asymptotic contribution to the coefficient $\beta$. 
Therefore, if $ q\vert\tau_M \vert \gg 1$, 
the positive frequency modes $\varphi_q^{M}$ of Eq.\ \Eqref{Mmode}
hardly differ from the asymptotic modes $ \varphi_{q}^{-\infty}$ since 
the magnitude of the corrections is controlled by 
some power of the small quantity $1/q\tau_M$.

Using \Eqref{t_M} and \Eqref{slowrolltau}, one finds
\begin{eqnarray}
  q\tau_M = M a(\tau_M) \tau_M 
  \simeq (1+\epsilon_1)\frac{M}{H(\tau_M)} =
   \frac{1+\epsilon_1}{\sigma_q} 
 \gg 1 \, .
\end{eqnarray}
Hence the norms of the Bogoliubov coefficients \Eqref{alphabeta} are
\begin{eqnarray} \label{normBogol}
  \vert \beta_q \vert^2 =  O\left(\sigma_q^{2p} \right) \, , \qquad
  \vert \alpha_q \vert^2 =  1 + \vert \beta_q \vert^2 \, ,
\end{eqnarray}
where $p \geq 1$ and the second equality follows from \Eqref{newmode} and 
unitarity.

When using the vacuum associated with the solutions of
Eq.\ \Eqref{Mmode}, one finds $p=3$ \cite{NPC1}. If one chooses instead
the solutions defined by $\left( i\partial_{\tau} - q
\right)\varphi_q^{M} = 0$, one obtains $p = 2$ \cite{MB03}, while those obeying 
$\left( i\partial_{\tau} - q  \right)(\varphi_q^{M}/a) = 0$
yield $p = 1$ \cite{Danielsson}.
This hierarchy is explained by the property that the corresponding states are 
instantaneous ground states at $\tau_M$
of Hamiltonians of decreasing degree of adiabaticity \cite{Veneziano}.
Hence, from a phenomenological point of view, the values of $p$ and $\sigma$
 should be conceived as free parameters characterizing 
the departure from the standard power spectrum.

Let us now consider the leading correction to the power spectrum in this
class of models. For $\sigma \ll 1$, it is always given by the second
term of \Eqref{newpower} which is linear in the coefficient $\beta_q$. 
The crucial point is that its phase is universally given by
\begin{eqnarray} \label{genericphase}
  \arg\left\{ \alpha_q \beta_q^* \left[\varphi_q^{-\infty}(q\tau \ll
      1)\right]^2  \right\} = 2 q\tau_M + O(1) \, .
\end{eqnarray} 
To derive this result, we first  
recall that the phase of the standard mode $\varphi_q^{-\infty}$ is calculated 
in the long wavelength limit, $q\tau \ll 1$,
while the phases of $\alpha_q$ and $\beta_q$ of  Eqs. \Eqref{alphabeta} 
are time independent. They depend parametrically on the $M$-crossing
time $\tau_M$ since the positive frequency modes $\varphi_q^M$ are
defined at that time. The result \Eqref{genericphase} is therefore governed by
the behavior of the modes in the two asymptotic regimes 
$q\tau \gg 1$ and $q\tau \ll 1$.

First, in the long wavelength limit, the phase
of the standard mode $\varphi_q^{-\infty}$ tends to a constant.
This can be seen from the asymptotic behavior of the
solution to \Eqref{modeeqn} in the slow roll approximation
\begin{eqnarray} \label{genericlonglambda}
  &&\varphi_q^{-\infty}  
  \sim \frac{-i \Gamma(\nu)}{\pi} \sqrt{-\tau}\left( \frac{-q\tau}{2} 
   \right)^{-\nu}\, ,
\end{eqnarray}
where $\nu = \frac{3}{2} + (f - 2)/3$ is real and depends on the slow
roll parameters $\epsilon_1$ and $\epsilon_2$. Hence
$\varphi_q^{-\infty}(q\tau \ll 1)$ makes an $M$-independent constant
contribution to the phase \Eqref{genericphase}.

Second, well inside the horizon the modes are close to Minkowski plane waves. 
Given their definition in Eq.\ (\ref{alphabeta}), the phase of 
$\alpha_q\beta_q^*$, i.e., the relative phase of $\alpha_q$ and $\beta_q$,
 is necessarily $2q\tau_M + O(1)$. 
To show this, it is sufficient to expand the solution $\varphi_q^{-\infty}$
in powers of $1/q\tau$,
\begin{eqnarray} \label{genericlonglambda}
  \varphi_q^{-\infty} = \frac{e^{-iq\tau}}{\sqrt{2q}} \left( 
  1 + \frac{b_1}{(q\tau)} + \frac{b_2}{(q\tau)^2} +
  O\left(\frac{1}{(q\tau)^3}  \right)    \right) \, , 
\end{eqnarray}
and notice that its time derivative can be factorized as
\begin{eqnarray} \label{expandphi'}
  i\partial_{\tau} \varphi_q^{-\infty} = q \varphi_q^{-\infty} \left( 
  1 - \frac{ib_1}{(q\tau)^2} + O\left(\frac{1}{(q\tau)^3}  \right)
\right)  \, . 
\end{eqnarray}

We consider the class of positive frequency modes  $\varphi_q^{M}$
that satisfy at time $\tau_M$
\begin{eqnarray} \label{defphi^M}
  (i\partial_{\tau}  - \Omega_q )\, 
  \left( \frac{\varphi_q^{M}}{a^n} \right) = 0 \, ,
\end{eqnarray}
where $n$ is a real number, and $\Omega_q$ is a function of the form
\begin{eqnarray} \label{Omega}
  \Omega_q = q \left( 1 + A\sigma_q^{m} + O\left(
      \sigma_q^{m+1}\right)\right)\, . 
\end{eqnarray}
For specific examples, see the paragraph below \Eqref{normBogol}.

Using Eq.\ \Eqref{expandphi'} evaluated at $\tau_M$ and
Eq.\ \Eqref{defphi^M}, we get  
\begin{eqnarray} \label{genericalpha}
   \alpha_q &=& \varphi_q^{M} \varphi_q^{-\infty \, *}\left\{ 
  \Omega_q+ i n aH(\tau_M) + q\left[ 1 - i\frac{b_1}{(q\tau_M)^2}
  + O\left(\frac{1}{(q\tau_M)^3}  \right)  \right]
  \right\} \, ,
\nonumber \\
 &=& 1 + \frac{A\sigma_q^{m}- in\sigma_q + 
     ib_1\sigma_q^2 }{2} 
     + O\left(\sigma_q^3 \right)\, ,
\nonumber \\
  &=& 1 + O\left( \sigma_q^{p'} \right)\, ,
\end{eqnarray} 
where $p'$ obeys $p \leq 2p' \leq 2p$ with $p$ defined in (\ref{normBogol}). 
It is related to the parameters $n,\,m,\, b_1$ as follows.
If $n \neq 0$, it is equal to ${\rm Min}(1,m)$, 
if $n=0$ and $b_1 \neq 0$ it is equal to  ${\rm Min}(2,m)$, and so on
(recall that $\sigma \ll 1$). In the second line, we used the liberty
to choose the overall phase of $\varphi_q^M$ 
to set it equal to that of $\varphi_q^{-\infty}$ at $\tau_M$
(see the remark below Eq.\ \Eqref{newpower}). 
In the last line we used Eq.\ \Eqref{Omega}. 
The coefficient $\beta$ is evaluated along the same lines:
\begin{eqnarray}\label{genericbeta}
   \beta_q &=& - \varphi_q^{M} \varphi_q^{-\infty}\left[ 
  \Omega_q + i n \sigma_q - q\left( 1 -  i b_1 \sigma_q^2 + 
  O\left( {\sigma_q^3}  \right)   \right)  \right] \, ,
\nonumber \\
  &=& \tilde B \, \sigma_q^{p} \,   e^{-2  iq\tau_M} 
  \, ,
\end{eqnarray}
where $\tilde B$ is a complex number 
whose phase originates from the correction terms $1/(q\tau_M)^{k> p}$ and
which is therefore practically independent of $q$ and $M$.
The result \Eqref{genericphase} follows from the combination of 
Eqs. \Eqref{genericlonglambda}, \Eqref{genericalpha} and \Eqref{genericbeta}.

The phase shift \Eqref{genericphase}
has a simple physical interpretation: 
it is (twice) the phase accumulated from the creation time $\tau_M$
until some time long after horizon exit when the power spectrum is evaluated
and the phase of the standard modes freezes out.  
Compared with the Bunch-Davis vacuum, the state $\ket{\Psi_M}$ 
contains pairs of quanta created at the time $\tau_M(q)$. 
The fact that quanta are created in pairs is reflected in the 
factor of two of the phase of the coefficient $\beta$ \cite{mapa}. 
Notice also that
the phase shift $q \tau_M$ is approximatively the redshift factor between 
the two scales $M$ and $H_q$, i.e., 
between the creation time and the time of horizon exit. 
This can be seen from
\begin{eqnarray}
  q\tau_M = \frac{\tau_M}{\tau_{H}} \simeq \frac{(aH)_{\tau_H}}{(aH)_{\tau_M}} 
 \simeq \frac{a(\tau_{H})}{a(\tau_M)}  \, ,
\end{eqnarray}
where we used that in the slow roll approximation 
$H(\tau) = H_0\left[1 + \epsilon_1 \ln(\tau/\tau_0) \right]$, so that  
$H$ changes by a constant of order unity during the $\ln\left(\sigma_q^{-1}\right)$
e-folds from $\tau_M$ to $\tau_{H}$.

In summary, the correction to the power spectrum is the sum of
two terms.
A square, which we call the steady term, and an interference term that 
depends on $M/H_q$ through its phase, called the oscillatory correction.
The steady correction is subleading and given by
\begin{eqnarray} \label{steady} 
  \vert \beta_q \vert^2 = O\left( \sigma_q^{2p} \right)\, ,
\end{eqnarray}
while the oscillatory correction is the leading
correction and given by the real part of
\begin{eqnarray} \label{phase}
 \frac{\beta_q^*}{\alpha_q^*} \frac{\left( \varphi_q^{-\infty}  \right)^2}{
 \vert \varphi_q^{-\infty}  \vert^2} = \tilde B^* 
  \, 
  \sigma_q^p \left(1+   O\left(  \sigma_q
   \right) 
  \right) \, e^{i2q \tau_M } 
   \, . 
\end{eqnarray} 
This term produces a modulation of the power spectrum.
Indeed, using \Eqref{sigmaq} and
$H(\tau) = H_0\left[1 + \epsilon_1 \ln(\tau/\tau_0) \right]$
which is valid  in the slow-roll approximation, one gets 
\begin{eqnarray} \label{expandx_q} 
 \sigma_q &=& \sigma_0 \, \left[1 - \epsilon_1 \ln\left( \frac{q}{q_0}   
 \right) \right] \, ,
 \nonumber\\
 q \tau_M 
 &=& \frac{1}{\sigma_0} \left[1+ \epsilon_1 + \epsilon_1 \ln\left( \frac{q}{q_0}   
 \right) \right] \, .
\end{eqnarray} 
From this we deduce that the period of the oscillations of 
the power spectra in $q$-space is given by 
\begin{eqnarray}
  \Delta \ln q  = \frac{\pi \sigma_0}{\epsilon_1}\, .
\end{eqnarray} 
It is linear in $\sigma \ll 1$,  
but also inversely proportional to the first slow roll parameter 
$\epsilon_1 \ll 1 $. Hence, its magnitude is determined by a
competition between the inflationary background 
evolution and the value of $H/M$.

Before determining the impact of averaging over fluctuations of the $UV$ scale $M$, 
we present a general expression for possible deviations of the primordial
power spectrum.

\section{A generalized Ansatz}

\label{sec:gener-other-class}

In the previous section, 
we considered the particular class of models where a prescription for 
the vacuum state is given at some finite time.
These are parameterized by only one dimensionless quantity, 
namely $\sigma_q$, which controls both the phase and the amplitude 
of the correction terms.
There is, however, no reason to assume this 
will be always the case when dealing with a fundamental theory. 
We therefore consider the more general expression of modifications
\begin{eqnarray} \label{paramP}
  {\cal P}(q) &=& 
  {\cal P}_{-\infty}(q)\left\{ 1  
  +  B_{q_0} \left( \frac{q}{q_0}  \right)^{\beta}  \cos\left[
 2\delta \ln\left( \frac{q}{q_0}  \right) + \psi   
 \right] + C_{q_0} \left( \frac{q}{q_0}  \right)^{\gamma}  \right\} \, .
\end{eqnarray}
where $q_0$ is a fiducial scale.

In writing this Ansatz, we assume that the deviations from new physics
are constrained to mild departures from scale invariance.
In other words, we do not account for phenomena that induce either 
sharp features (i.e., in the form of $\delta_{\rm Dirac}(q- q_0)$)
or rapidly oscillatory behavior such as $\cos(q/q_0)$. 
What motivates our choice is the fact that standard physics 
is nearly scale invariant, in that the deviations of the power spectrum from
scale invariance only come from the background geometry through 
logarithmic derivatives of $H_q$. This follows from the
near-stationarity of the amplification process of successive modes
with increasing conformal scale $q$.
Hence, under the assumption that the new physics preserves this stationarity,
the $q$-dependent corrections are still governed  by 
$\epsilon_1 \ln q$ as it was the case in  
Eqs. (\ref{expandx_q}).\footnote{Anticipating 
Section \ref{sec:geometric-average}, it is interesting
to notice that sharp modifications of the
primordial power spectrum 
(in the sense that they vary much more rapidly than $\ln q$) 
are strongly broadened and damped 
by the geometric projection involved in computing the
angular power spectrum of the CMB \cite{featuresinCMB}.
Hence, from the point of view of confronting CMB data
they need not be considered.}

The modified power spectrum \Eqref{paramP} is described by $6$ new 
parameters. The terms proportional to $B$ and $C$ represent the oscillatory
and the steady corrections, respectively. 
The model of Section \ref{sec:steady-non-steady} is contained in
\Eqref{paramP}, with the special values 
$\psi =  2(1+\epsilon_1)/\sigma_0$  
and $\delta = \epsilon_1 /\sigma_0$ as seen from 
\Eqref{steady}-\Eqref{expandx_q}. 
The other coefficients can be obtained by Taylor expanding in 
powers of $\ln(q/q_0)$ around the fiducial point $q_0$,
see for instance \cite{MB03} for detailed expressions.
The Ansatz \Eqref{paramP} also includes extensions of these models  
which allow for a so-called $\alpha$-vacuum in place of the adiabatic vacuum 
(the transformation \Eqref{newmode} is combined with a second 
Bogoliubov transformation). 
These power spectra are considered, for 
instance, in  \cite{MR} and \cite{MB03} and are described 
by three independent parameters.

Notice also that the parameterization \Eqref{paramP} allows for 
a combination 
of various subleading corrections (possibly characterized by several scales) 
to the standard slow-roll power spectra, but 
not necessarily of a high energy origin.
This is particularly clear for the steady term whose Taylor expansion is
\begin{eqnarray} \label{expandC}
  Cq^{\gamma} = 1 + \gamma C \ln\left(\frac{q}{q_0}\right) + 
 \frac{1}{2}\gamma^2 C \ln^2\left(\frac{q}{q_0}\right) + ...
\end{eqnarray}
A calculation of the power spectrum beyond the slow roll approximation yields
a result of the same form \cite{beyondSL}
\begin{eqnarray}
  P_{-\infty}^S = \frac{H^2}{\pi \epsilon_1 M_{\rm Pl}^2} \left[
  a_0 +  a_1 \ln\left( \frac{q}{q_*}  \right) + a_2\ln^2\left( \frac{q}{q_*}  \right)
 + ...
 \right]
\end{eqnarray}
where the coefficients $a_i$ depend now on the parameters
$\epsilon_n$ and are of order $O(\epsilon_n^i)$.
A similar expansion holds for the power spectrum of primordial gravitational
waves.

The corrections from matter loops make another contribution to the
coefficient $\gamma$  \cite{Weinberg}:
\begin{eqnarray}
  P_{-\infty}^S = \frac{H^2}{\pi \epsilon_1 M_{\rm Pl}^2} \left[ 1+ 
  b_{1-\rm loop} GH^2 \epsilon_1(t_H) \ln\left( \frac{q}{\mu}  \right) + 
  O(G^2H^4 \epsilon_1^2) \right] 
\end{eqnarray}
where $b_{1-\rm loop}$ is a numerical factor and $\mu$ the
renormalisation scale.

Consequently, high energy
corrections of the type \Eqref{expandC} may be hard 
to disentangle from non-trivial standard physics effects, such as the
slow rolling background or loop contributions.

\section{Stochastic averaging}

\label{sec:stochastic-averaging}

To model the consequences of $UV$ 
geometric fluctuations which might arise in quantum gravity,
we treat $M$ as a fluctuating variable and calculate the
power spectra after taking the average over its fluctuations. 
More precisely, we adopt the simplest description  
by assuming that $M$ is a Gaussian variable characterized by a
mean $\bar M$ and a spread $\Sigma$
\begin{eqnarray}
  \av{M} = \bar M \, , \qquad 
 \av{(M-\bar M)^2}^{1/2} = \Sigma \, .
\end{eqnarray}

We also assume that the spread is much smaller than the mean
\begin{eqnarray} \label{cond1}
   \Sigma \ll \bar M \, , 
\end{eqnarray}
as in the Breit-Wigner description of long living atomic states. 
This requirement implies that the induced spread  
of the cosmological time $t_M$ defined at Eq.\ \Eqref{t_M} 
is much smaller than the Hubble time $1/H$ 
evaluated at the mean time $\bar t$ defined
by $q = \bar M a(\bar t)$. Indeed, 
by differentiation of the relation $q = M a(t_M) $ at fixed $q$, we have
\begin{eqnarray} \label{deltaN}
  \delta \ln a_M =  H_M \delta t_M = \frac{\Sigma}{\bar M}  \ll 1 \, .
\end{eqnarray} 
Similarly, the parameter $\sigma_q$  now also exhibits some 
spread $\delta \sigma$. Again, the ratio of this variance 
over the mean  $\bar \sigma_q$
obeys $\delta \sigma /\bar\sigma = \Sigma /\bar M$.

To characterize the effects of the fluctuations,
it will be convenient to parameterize the spread of $M$ by a
power $n$ defined as follows:
\begin{eqnarray}
  \Sigma_n \equiv H_M \left( \frac{\bar M}{H_M} \right)^n \, .
\end{eqnarray}
The condition \Eqref{cond1} now reads
\begin{eqnarray} \label{constraintn}
   n- 1 < \frac{3}{\ln\left(\bar M /H \right)} \, ,
\end{eqnarray}
where we have adopted the convention $e^3 \gg 1$.

We now take the ensemble average of \Eqref{newpower}. 
Let us consider each term separately. The steady correction
is basically unchanged because the norms of Bogoliubov coefficients
are slowly varying functions of $M$. Hence the mean value 
$\av{\vert  \beta_q \vert^2}$ is well approximated by its former expression
(\ref{steady}) evaluated with the mean quantity $\bar \sigma_q$, that is 
\begin{eqnarray} \label{avsteady}
   \av{\vert \beta_q \vert^2} = O\left( \bar \sigma_q^{2p}   \right) \, .
\end{eqnarray}

On the contrary, averaging over the fluctuations of $M$ in
the oscillatory term \Eqref{phase} has a dramatic effect.
Indeed, its mean value is damped by an exponential factor 
\begin{eqnarray} \label{statdamping}
  \av{\sigma_q^p \, e^{i2/\sigma_q}} &=& \bar \sigma_q^p \, e^{i2/\bar \sigma_q}
 \times 
  \exp\left[ - 4 \frac{H_M^2}{H_q^2} \, \left( \frac{\bar M}{H_M} 
  \right)^{2n}\right] \,  .
\end{eqnarray}
To evaluate the prefactor in the exponential, we use again the slow roll
approximation wherein
$H(\tau_M) = H_q\left[1 + \epsilon_1 \ln(\tau_M/\tau_{\rm ex}) \right] \simeq 
 H_q \left[1 - \epsilon_1 \ln(\sigma_q) \right] \simeq H_q$. 
For instance, with $\epsilon_1 = 10^{-2}$ and $\sigma_q = 10^{-4}$, we have
$H_M = 1.1 \times H_q$. In any case, one has $H_M / H_q > 1$, something 
which increases the damping factor. For simplicity, we will take it equal to one.

Since one has $\bar M/H = 1/\bar \sigma \gg 1$, 
unless $n<0$, that is unless $\Sigma < H_M$,the oscillatory
term is exponentially suppressed by a large quantity.
To appreciate the importance of this effect, it is instructive to compute 
the value of $n$ such that the damping of the oscillatory term 
reduces it to the subleading correction represented by the steady term. 
The averaged values of the oscillatory and steady 
corrections are of the same order for $n$ given by
\begin{eqnarray} \label{theN}
  n_{\rm eq} = 
  \frac{\ln\left( \frac{p}{4} \ln\frac{\bar M}{H_M} \right)}
 {2\ln\frac{\bar M}{H_M}}\, .
\end{eqnarray}
For instance, if we choose $\bar M / H_M = 10^{4}$ and $p = 1$, we find 
$n_{\rm eq} \simeq 5/100$. 
This is perfectly
compatible with the constraint \Eqref{constraintn} 
which reads for these numerical values $n<1.32$.
For $n> n_{\rm eq}$, the oscillatory term is so damped by the ensemble
average that it becomes smaller that the steady correction \Eqref{avsteady}
which thus provides the new leading deviation.

This conclusion has been reached for the models of
Sec. \ref{sec:steady-non-steady},  
but it can be readily generalized to deviations of the power spectra 
parameterized by \Eqref{paramP}. By assumption, the function $C(M)$
weighing the steady term is a slowly varying function of $M$, so that 
in a first approximation it may be replaced by its value 
at the mean $\bar C = C(\bar M)$, as
in \Eqref{avsteady}. Instead, the oscillatory term proportional to $B$ 
must be treated similarly to \Eqref{statdamping}. That is, whenever
the functions $\delta(M)$ or $\psi(M)$ 
change significantly over an interval $\Sigma$, 
the ensemble average of the oscillatory term will be damped 
\begin{eqnarray} \label{avcos}
  \av{\cos\left[
 2\delta \ln\left( \frac{q}{q_0}  \right) + \psi   
 \right]} =  \cos\left[
 2\bar \delta \ln\left( \frac{q}{q_0}  \right) + \bar \psi   
 \right] \times \exp\left( - K \left(\frac{\bar M}{H_M} \right)^{2n'} 
 \right)\, .
\end{eqnarray}
where $K$ is a constant. 
The value of $n'$ is determined by the fastest oscillating term. 
For instance, if $\psi(M)$ is again linear in $M$, one still finds $n'=n$.

In general the equality \Eqref{theN} will be 
replaced by
\begin{eqnarray} \label{n_eq'}
  n_{\rm eq}' = 
  \frac{1}{2\ln\left( \bar \sigma^{-1}  \right)} \ln\left[ \frac{\ln(B/C)}{K}
   \right]\, .
\end{eqnarray} 
Hence, provided $K$ is not too small, 
the power $n'$ may be rather small while the 
oscillatory term can still be severely suppressed.

In conclusion, unless $\Sigma < H_M$, 
the oscillatory  deviations of the power spectrum are strongly reduced
and become subleading corrections to the averaged power spectra.

\section{Geometric averaging}

\label{sec:geometric-average}

In this section, we compare the high energy averaging 
of the deviations of the primordial power spectrum 
to the geometric averaging involved   
in computing the two-dimensional angular power spectrum. 
The contribution from scalar perturbations to the 
multipole $l$ of the temperature anisotropies  
can be written as \cite{CMBslowWeinberg}
\begin{eqnarray} \label{theCl}
  C^S_l &=& \frac{2}{\pi} \, \int_0^{+\infty}\!\!\frac{dq}{q} \, {\cal P}_{S}(q) 
 \left[ {\cal T}_{\rm int}(q) j_l(q d_A)  
 + {\cal T}_{\rm v}(q) j_l'(q d_A) \right]^2\, ,
\end{eqnarray}
which nicely separates the contributions of the physics and the geometry. First, 
the spherical Bessel function $j_l$ acts as a projector on the celestial sphere, 
where $d_A$ is the angular diameter distance 
of the last scattering surface. For flat spatial sections, it is 
given by the lapse of conformal time since last scattering, i.e. 
$d_A = \tau_0 - \tau_{lss}$.

Second, the curvature power spectrum ${\cal P}_{S}(q)$ seeds the various 
matter and radiation density perturbations, the 
evolution of which is encoded in the transfer functions ${\cal T}$.
${\cal T}_{\rm int}$ 
essentially describes intrinsic temperature fluctuations and the 
Sachs-Wolfe effect, while ${\cal T}_{\rm v}$ is due to the Doppler effect.

On large angular scales, i.e. for $l \, d_H/d_A \ll 1$, where $d_H$ is the size 
of the acoustic horizon (in practice $l \ll 100$), 
one can use the following asymptotic expressions of the form factor
\begin{eqnarray} \label{ITformfactors}
  {\cal T}_{\rm int}(q) = 1 + O\left( q^2 \right) 
  \, , \qquad 
  {\cal T}_{\rm v}(q) = O\left( q \right)\, ,
\end{eqnarray}
In this case, as noticed in \cite{MR}, 
the integral \Eqref{theCl} actually performs a geometric average over 
the fine-structure of the primordial power spectrum.
More precisely, for the power spectrum 
given by \Eqref{paramP}, the oscillatory  
deviations are damped by a power of the frequency of 
superimposed oscillations $\epsilon_1  / \sigma_0$ while
the steady correction is not.

In the limit \Eqref{ITformfactors},
the integral \Eqref{theCl} can be done explicitly with the change
of variables $s=qd_A$ and with the help of
\begin{eqnarray}
  I_l(m) \equiv \int_0^{+\infty}\!\!\frac{ds}{s} \, s^{m} \,  j_l^2(s) = 
  2^{m-3} \pi \frac{\Gamma(2-m) \Gamma\left( l +\frac{m}{2} \right)}{
 \Gamma^2\left(\frac{3-m}{2} \right) \Gamma\left(l+2 -\frac{m}{2} \right)}\, ,
\end{eqnarray}
applied for the values $m=\gamma$ and 
$m=\beta + i 2\delta$ for the steady and oscillatory corrections
respectively.  

To simplify the expressions we set $2\epsilon_1 + \epsilon_2 = 0$
in ${\cal P}_{-\infty}^{\zeta}$
(flat power spectrum). 
In this case, we find
\begin{eqnarray} \label{modifC_l}
  C^S_l &=& C_l^0 \, \left\{ 1  + \,B_{q_0} \, 
  (q_0 d_A)^{-\beta} 
 \, {\rm Re} \left[ e^{i(\psi - 2\delta\ln(q_0 d_A))}  \,   
{I_l(\beta+i2\delta)\over I_l(0)}  \right] \right.
 \nonumber \\
 && \left. \quad \quad \,\,\, + \,
 C_{q_0} \,  (q_0d_A)^{-\gamma}\, { I_l(\gamma)\over I_l(0)}
 \right\} \, ,
\end{eqnarray}
where $C_l^0 \propto I_l(0) \propto 1/l(l+1)$
is the unperturbed power spectrum. 
As in Eq.\ (\ref{paramP}), $B_{q_0}$ and $C_{q_0}$ weigh 
the oscillatory and steady corrections respectively.

 For the steady correction, if  $\gamma$ is not too large,   
is of order unity for all $l$.  
Hence, as one might have expected, 
the projected amplitude of the (relative) steady correction does
not significantly differ from its original amplitude in 
the power spectrum (\ref{paramP}).

For the oscillatory corrections, we again consider 
the case where the oscillations have a high frequency.
In this case, the parameter $\delta \gg 1$ in \Eqref{paramP}.
(We recall that in the models of section II, $\delta = M \epsilon_1 /H_M \gg 1$.)   
Using the Stirling formula to evaluate $I_l(\beta+i2\delta)$ yields  
\begin{eqnarray} \label{LOCl}
  \frac{\Delta_{\rm oscill.} C_l^S}{C_l^S} \propto - 
  \frac{B_{q_0}}{ (q_0 d_A)^{\beta}
  \delta^{5/2}}
  \cos\left[
  2\delta\ln\left(\frac{\delta}{a_0Md_A}  \right)  + 
  \pi l + \psi - \frac{\pi}{4}   
  \right]\, ,
\end{eqnarray}
Instead of finding an exponential damping as in Section
\ref{sec:stochastic-averaging}, we obtain a power law suppression governed by $5/2$,
in agreement with Eq.\ (13) in \cite{MR}.   
Therefore, the oscillatory deviations of the primordial spectrum
 provides the leading correction to multipoles only if 
\begin{eqnarray} \label{condleading}
  \frac{B_{q_0}}{C_{q_0}} (q_0 d_A)^{\gamma - \beta} 
  > \delta^{5/2} \, .
\end{eqnarray} 
One can clearly see that the geometric projection introduces a 
preferred conformal scale through the angular diameter distance $d_A$,
in contrast with the scale independence of the 
damping of stochastic origin, see \Eqref{avcos}.

Finally, to confront deviations originating from
new high energy physics to observable data, 
it is also necessary to evaluate $n_G'$,
the value of $n'$ of Eq.\ \Eqref{avcos} such that the damping 
factor
from the new physics equals 
that from the geometric average in
\Eqref{LOCl}. Their equality means 
\begin{eqnarray}
   \exp\left[ - \frac{K}{\bar \sigma^{2n_G'}}    \right] = 
  \delta^{-5/2} \, (q_0 d_A)^{ - \beta} 
\end{eqnarray}
where $K$ is a constant of order unity. In turn, this implies 
\begin{eqnarray}
  n_G' = \frac{1}{2\ln\left( \bar \sigma^{-1}  \right)} \, 
  \ln\left(\frac{5}{2K} \ln \delta - \frac{ \beta}{2K} \ln(q_0d_A)
  \right) \, , 
\end{eqnarray}
which is the same equation as \Eqref{n_eq'} with the substitution 
$B/C \mapsto \delta^{5/2} (q_0 d_A)^{- \beta}$. 
As a consequence, unless $n' < n_G'$, the oscillatory correction term
is damped by a factor larger than $\delta^{-5/2}$.

\section{Conclusion}

\label{sec:discussion}

The geometric and stochastic averages are cumulative, in the sense that the 
oscillatory correction receives a second 
 damping factor $\delta^{-5/2}$ from the
integral over the wavenumbers. It multiplies the first damping factor 
$\exp\left( - K \bar \sigma_q^{-2n'}  \right)$ from the average over $M$.

However, these two averaging procedures differ in the following 
important way. 
The stochastic average is (almost) scale independent, in the sense that 
the oscillatory correction to the power is damped by the same factor 
$\exp\left( - K \sigma_0^{-2n'}  \right)$ independently of the wavenumber (almost here 
must be understood in the same way as the unperturbed power spectrum is almost scale 
invariant, that is with a slow logarithmic dependence in $q$).

On the other hand, the geometric average considered in the previous section 
is only valid for large angular scales, for which the form factors 
can be approximated by constants. 
For smaller angular scales, the ${\cal T}$'s are oscillatory functions of $q$ 
with a frequency equal to the size of the acoustic horizon \cite{CMBslowWeinberg}. 
They therefore interfere 
with the superimposed oscillations to the primordial power spectrum and 
produce potentially observable oscillations in the angular power spectrum \cite{MR}.
In other words, the geometric averaging depends on the angular scale $l$ while
the stochastic averaging does not.
It is therefore possible to distinguish them in principle.

In brief, two cases leading to different lessons can be found. 
If the detection of the superimposed oscillations in the CMB data are 
confirmed, this would constitute a very strong 
constraint on the width $\Sigma$ of the $UV$ scale $M$.
If instead the $UV$ damping of the oscillatory term 
is so strong that the steady term becomes
the leading correction, no further damping would be introduced
by the geometric averaging, and the corrections would be 
proportional to $\vert \beta \vert^2 \propto (\sigma^{p})^2$. This may
well be too small to make them observable.

Finally, as noticed in Section \ref{sec:gener-other-class}, 
the steady corrections to the power spectra receive 
contributions from various physical effects. Lifting the induced
degeneracy which 
impedes the access to information about Quantum Gravity is a challenge
for future work.

\acknowledgements

The work of DC
and JCN was supported by the Alfried Krupp Prize for Young 
University Teachers of the Alfried Krupp von Bohlen und Halbach
Foundation.

\end{document}